# Demonstration of ultra-broadband single-mode and single-polarization operation in *T*-Guides

JEFF CHILES,[1] AND SASAN FATHPOUR,[1,2,*]

[1] CREOL, The College of Optics and Photonics, University of Central Florida, Orlando, FL 32816
[2] Department of Electrical Engineering and Computer Science, University of Central Florida
*Corresponding author: fathpour@creol.ucf.edu

**Silica-based anchored-membrane waveguides (*T*-Guides) are fabricated and characterized from the visible to infrared with streak imaging. It is numerically shown that the *T*-Guides can have wideband single-mode and single-polarization (SMSP) properties over a span of 2.6 octaves. Experimentally, a polarization-dependent loss difference of up to 90 dB/cm is measured between orthogonal polarizations, and a record SMSP window of >1.27 octaves is observed, limited only by the available measurement equipment. These measurements make a strong case for *T*-Guides for SMSP photonics, particularly on high-index materials such as our previous demonstration on silicon.**

The modal properties of waveguides can strongly affect their performance in many important application sectors. In telecommunications, modal dispersion leads to pulse distortion, and coupling into higher-order modes during bends results in power loss. In the context of supercontinuum generation, operating with multimode waveguides can result in poor beam quality of the output, which limits the brightness achievable. It is also desirable to achieve a highly polarized output, without simply rejecting orthogonally-polarized light. One means of averting these issues is through single-mode and single-polarization (SMSP) waveguides. SMSP waveguides should be designed to maximize the SMSP "window" over which one mode and one polarization can propagate with sufficiently low loss. Any other mode or polarization is either completely unsupported, or made to be strongly attenuated. The specific means of achieving this may vary, but in general, asymmetric geometries or material anisotropy are employed [1].

To date, SMSP waveguiding has been primarily investigated in fibers [2-6]. However, the bandwidth of fabricated fibers to date has reached only 0.23 octaves [6], and for most designs, the induced loss for the undesired polarization is usually measured in the meter or kilometer length-scale. For compact and low-cost photonics, an integrated SMSP solution with much stronger attenuation for the undesired polarization and higher-order modes is desirable.

Recently, we have proposed a novel geometry known as "anchored membrane" waveguides (or the *T*-Guide) [7] and demonstrated it on an all-Si platform [8]. The variation on silica ($SiO_2$), as well as the simplified fabrication method on silicon substrates, are depicted in Fig. 1. It is numerically and experimentally shown in this work that these silica *T*-Guides possess SMSP windows significantly larger than the previous reports in integrated waveguides and optical fibers.

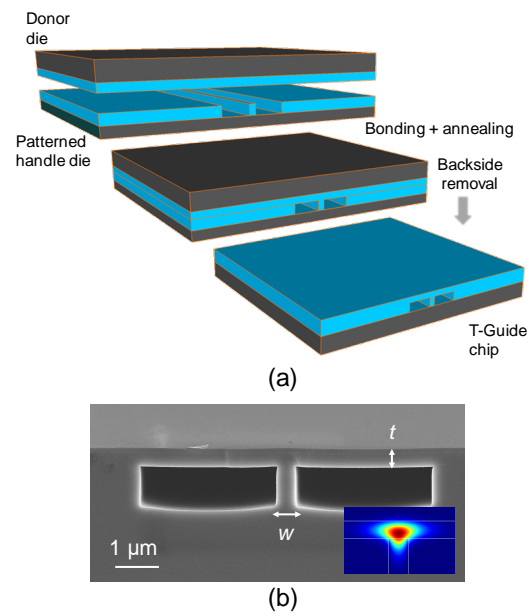

Fig.1. (a) Simplified fabrication method for *T*-Guides; (b) SEM cross section of a fabricated silica *T*-Guide on silicon substrates (inset: simulated intensity profile of the mode).

It consists of a membrane (or "slab") bonded over two trenches, separated by a narrow post, which acts to confine the optical mode at the junction of the "T." Its asymmetric and semi-infinite geometry enable extremely broad SMSP windows [7]. Only the fundamental transverse-electric mode is permitted for most designs and wavelength regions. Silicon *T*-Guide designs were theoretically investigated and shown to exhibit up to 2.75 octave-wide SMSP windows [7], spanning nearly the entire material transparency window of silicon [9].

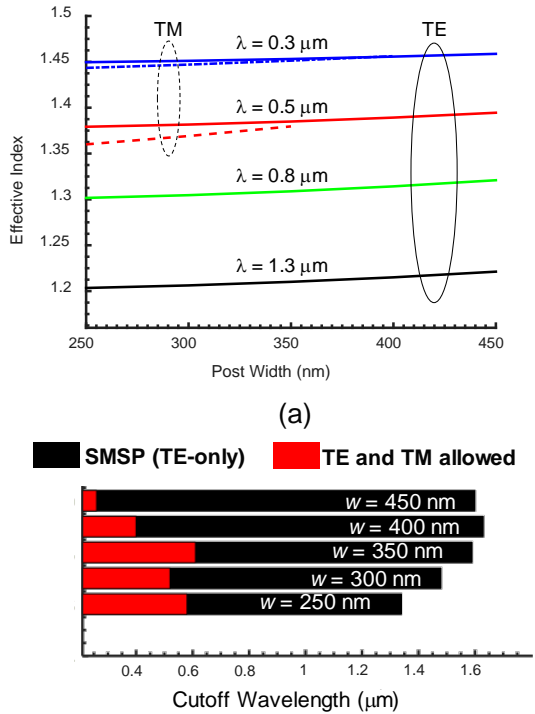

Fig. 2. (a) *T*-Guide post width vs. modal effective index for different wavelengths. Dashed lines are TM modes and solid lines are TE modes; (b) Transmission windows for the TE and TM guided modes, with black corresponding to SMSP operation and red to bi-polarized operation.

Furthermore, the geometry provides superior mechanical stability, thanks to its post feature, while allowing compact dimensions due to the large index contrast achievable between the core material and the air cladding. In this work, we experimentally demonstrate that *T*-Guides can exhibit ultra-broadband SMSP operation, as predicted.

We have recently demonstrated Si *T*-Guides in the mid-infrared [8]. However, in order to measure their SMSP properties, it is desirable to implement *T*-Guides such that accurate loss measurements can be made over a very broad bandwidth. Furthermore, since the magnitude of losses for the "rejected" polarization is expected to be in the order of > 50 dB/cm for much of the spectrum, the measurement setup must be capable of resolving this value over very short distances. This loss is too high for resonator-based loss measurement, or for cut-back measurements. An alternative can be found in top-view streak imaging [10], in which a digital image is captured from the top view of light scattering from the top surface of a waveguide. The intensity of scattered light is fitted to an exponential curve, allowing accurate loss measurements over a wide range, limited only by the detector sensitivity and the magnification optics. Therefore, they should be implemented in a material transparent in the visible to near-IR range where suitable cameras and optics are readily obtained. To this end, we chose to explore *T*-Guides constructed of silica, rather than silicon [9]. It is noted that silica exhibits wideband material transparency [11] and can be processed using techniques similar to silicon.

First, the theoretical SMSP operation of silica-type waveguides are considered, in order to understand what can be expected of various designs. Accordingly, eigenmode simulations of silica *T*-Guide geometries were conducted in COMSOL™ over a wide spectrum. Scattering boundary conditions were used, and a cutoff condition of 10 dB/cm from leakage losses was employed. Generally, the cutoff is chosen to correspond to the length scale of interest; in fibers, 1 dB/m has been used as a cutoff in the past [3]. The SMSP window is thus determined by the spectrum in which TM losses are above this limit, and transverse electric (TE) losses are below this limit (in an asymmetric structure, the TE mode will cut off when a sufficiently long wavelength is reached). The simulated structure in this case used a membrane width of 5 µm, a slab thickness of $t$ = 350 nm, a post height of 2.5 µm, and post widths, $w$, varying from 250 to 450 nm.

The effective index for transverse magnetic (TM) and TE modes are plotted for varying post widths in Fig. 2(a), and the corresponding window of operation for the mode for each post width and polarization is shown in Fig. 2(b). The observed results are similar to those observed in the theoretical investigations of silicon *T*-Guides [7], where both polarizations are guided only at short wavelengths, and increasing the post width beyond the slab thickness results in broader SMSP windows. This occurs because the TM mode begins to leak to the substrate through the post as it is widened. An SMSP window from 260 to 1600 nm (2.6 octaves) is possible for a post width of 450 nm with silica *T*-Guides, but smaller post widths also produce extremely broad SMSP operation. These designs thus provide a starting point for the experiment.

Silica *T*-Guides were fabricated according to the flow in Fig. 1(a). A thermally oxidized silicon die was patterned with trenches to form isolated posts between 250 – 400 nm wide. It was bonded to a thermally oxidized piece of silicon (the "donor" die) approximately 3 x 5 cm² and annealed at 1000°C for 1 hour. The backside of the donor die was removed by dry etching, followed by wet-etching with tetramethylammounium hydroxide to expose the silica membrane without damage. The thickness of the transferred membrane was adjusted from 300 to 350 nm using plasma-enhanced chemical vapor deposition. These processing steps resemble our approach for demonstration of all-Si suspended-membrane waveguides [12] and, of course, the aforementioned silicon *T*-Guides [8].

Cleaved waveguides were characterized to assess their SMSP window. Three different lasers were utilized: blue (diode, 405 nm), red (helium neon, 633 nm) and near-infrared (diode, 976 nm). Light was coupled into the *T*-Guides through an aspheric lens, and a top-view of scattered light from the waveguides was imaged onto a silicon camera module through a microscope (the infrared blocking filter was removed to permit infrared measurements). A linear polarizer was placed in front of each laser to image the TE and TM mode streaks separately for comparison. The corresponding streak images for *T*-Guides with widths of 250 and 300 nm are collected and shown in Figs. 3(a-b). It can be seen that for all wavelengths and for both widths, the TM mode is severely attenuated compared to the TE mode, as evidenced by the short length of the streak. Fitting the measured intensity along the streak to an exponential function, as in the example of Fig. 4, the propagation loss was examined for $w$ = 250, 300 and 350 nm. A high degree of particle scattering was observed during the measurements, evidenced by the bright dots in Figs. 3(a-b). It results from particles trapped in between the slab and the handle wafer during bonding. This discrete scattering (as opposed to the more continuous scattering observed due to nanoscale roughness in waveguide features) thus obscured most measurements of the TE mode propagation loss. Nevertheless, two TE mode loss cases could be accurately analyzed despite the scattering: $w$ = 300 nm at $\lambda$ = 405 nm with 12.3 ± 2 dB/cm, and $w$ = 250 nm at $\lambda$ = 976 nm with 19 ± 3 dB/cm. With higher-quality thermal oxide wafers, the loss can be greatly reduced in the future.

It is noted that we have recently measured losses of 1.75 dB/cm on silicon $T$-Guides at a wavelength of 3.6 μm [8]. However, as discussed before, it is very difficult to experimentally investigate their SMSP property over the wide predicted range of 1.2 - 8.1 μm [7].

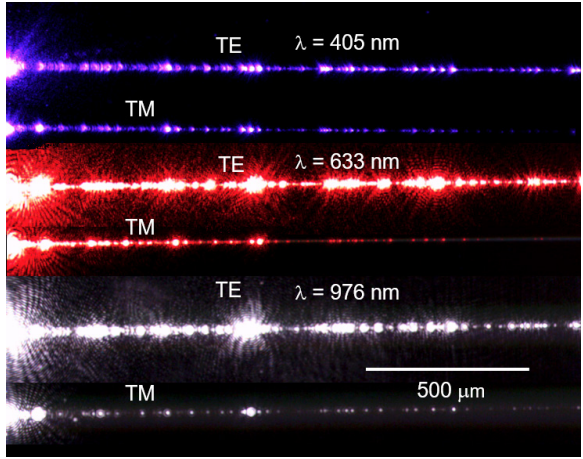

(a)

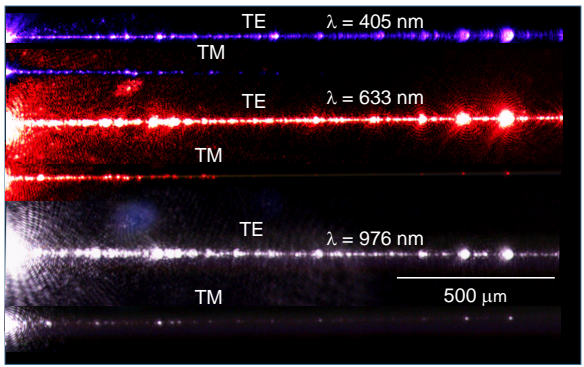

(b)

Fig. 3. Streak images of $T$-Guides: (a) $w$ = 250 nm; (b) $w$ = 300 nm. Single-polarization operation is observed by the much longer length of TE-mode streaks compared to those of the TM-mode.

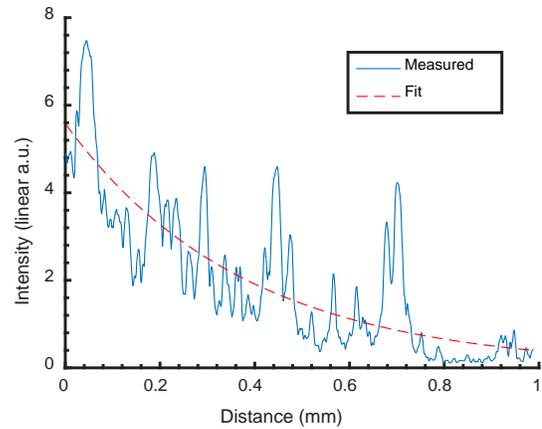

Fig. 4. Measured (solid) and fitted (dashed) intensity of scattered light for $w$ = 250 nm for the TM mode at $\lambda$ = 633 nm.

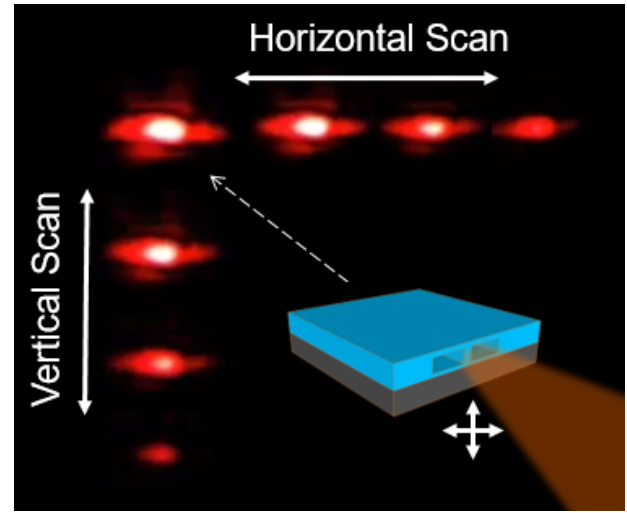

Fig. 5. Images of the optical mode collected from a $T$-Guide during horizontal and vertical misalignment of the focused input beam onto the facet. No changes to the spatial distribution were observed. The faint lines above and below the mode are a result of aberrations in the imaging optics used.

Next, the TM mode propagation losses (the rejected polarization) were examined. The case of $w$ = 300 nm showed the strongest attenuation for the TM polarization, a result consistent with the expected behavior of TM light leaking to the substrate through the wider post. At $\lambda$ = 405 nm, the corresponding TM mode propagation losses for each width were 49, 102 and 387 dB/cm, respectively. For $\lambda$ = 633 nm, the TM losses were 116, 181 and 234 dB/cm. The propagation loss for the TM mode at $\lambda$ = 976 nm was too large to measure accurately; however, since the largest measurable loss in this experiment was 387 dB/cm, the attenuation is expected to be even greater. The uncertainty in all loss measurements is estimated to be ±15%, which is sufficient for the large values observed. The earlier onset of single-polarization operation at shorter wavelengths here than that predicted by simulations can be explained by the slightly rounded top corners of the fabricated post (Fig. 1(b)), which significantly increase the rate of leakage through the post for the TM mode.

It should be noted that the modal cutoff condition of 10 dB/cm propagation loss used in the simulations cannot be straightforwardly applied to these measurements, since the performance of scattering boundary conditions is inherently different from substrate leakage in reality. Regardless, it is clear from the streak measurements that extremely high attenuation values are achieved for the TM mode in comparison to those of the TE mode, thus demonstrating strong single-polarization operation over a span of more than 1.27 octaves (405 to 976 nm) in the case of $w$ = 300 nm, limited by the lasers available for the measurement and the camera's spectral sensitivity. A maximum polarization-dependent loss (PDL) of 90 dB/cm was observed in the case of $\lambda$ = 405 nm for $w$ = 300 nm. The 1.27-octave SMSP window is significantly higher than the previous record of 0.23 octaves in optical fibers [6].

Although the single-mode behavior of *T*-Guides is derived from the same well-understood mechanism as for shallowly etched ridge waveguides [13], it is nevertheless desirable to provide an experimental confirmation of this. To this end, the tightly focused input beam at 633 nm wavelength was scanned both horizontally and vertically across a *T*-Guide facet (with $w$ = 300 nm) while the output mode profile was imaged with a camera (Fig. 5). The existence of higher-order modes would be evidenced by a change in the spatial distribution of the intensity during the scan. However, the mode retains its shape and only changes in absolute intensity, confirming that only one transverse mode is permitted in the *T*-Guides.

In conclusion, the single-mode and single-polarization (SMSP) properties of silica-based anchored-membrane (*T*-Guide) waveguides were numerically and experimentally investigated. Silica *T*-Guides were fabricated and characterized from the visible to infrared using streak imaging. Polarization-dependent losses of up to 90 dB/cm were measured, and an SMSP window > 1.27 octaves was experimentally observed, limited only by measurement equipment available. These results support the expected broadband SMSP behavior of these structures, and establish *T*-Guides as a promising candidate for high-quality on-chip nonlinear photonics.

**Funding.** National Science Foundation (NSF) CAREER (ECCS-1150672); Office of Naval Research (ONR) YIP (11296285).

**Acknowledgment**. We thank Prof. Kathleen Richardson's research group for the use of their Firefly OPO, and Prof. Axel Schülzgen's group for the use of their 976 nm laser diode source.